\documentclass[letter]{aa} 
\usepackage{graphicx}
\usepackage{txfonts}
\usepackage{natbib}
\bibpunct{(}{)}{;}{a}{}{,} % to follow the A&A style
\usepackage{hyperref}
\begin{document} 

\title{Transiting planet candidate from \textit{K2 \textup{with the longest period}}}

\author{H.A.C. Giles\inst{1}\fnmsep\thanks{Email: Helen.Giles@unige.ch}, H.P. Osborn\inst{2},
S. Blanco-Cuaresma\inst{3},
C. Lovis\inst{1},
D. Bayliss\inst{4},
P. Eggenberger\inst{1},
\\
A. Collier Cameron\inst{5},
M.H. Kristiansen\inst{6,7},
O. Turner\inst{1},
F. Bouchy\inst{1},
\and S. Udry\inst{1}}

\authorrunning{H.A.C. Giles et al.}

\institute{Observatoire de Gen\`{e}ve, Universit\'{e} de Gen\`{e}ve, Chemin des Maillettes 51, 1290 Versoix, Switzerland\\
\email{Helen.Giles@unige.ch}
\and
Aix Marseille Univ, CNRS, LAM, Laboratoire d’Astrophysique de Marseille, Marseille, France
\and
Harvard-Smithsonian Center for Astrophysics, 60 Garden Street, Cambridge, MA 02138, USA
\and
Department of Physics, University of Warwick, Gibbet Hill Road, Coventry, CV4 7AL, UK
\and
Centre for Exoplanet Science, SUPA, School of Physics and Astronomy, University of St Andrews, North Haugh, St Andrews KY16 9SS, UK
\and
DTU Space, National Space Institute, Technical University of Denmark, Elektrovej 327, DK-2800 Lyngby, Denmark
\and
Brorfelde Observatory, Observator Gyldenkernes Vej 7, DK-4340 T\o{}ll\o{}se, Denmark
}

\date{Received June 5, 2018; accepted June 22, 2018}

% \abstract{}{}{}{}{} 
% 5 {} token are mandatory

\abstract
{We present the transit and follow-up of a single transit event from Campaign 14 of \textit{K2}, EPIC248847494b, which has a duration of 54 hours and a 0.18\% depth.}
{Using photometric tools and conducting radial velocity follow-up, we vet and characterise this very strong candidate.}
{Owing to the long, unknown period, standard follow-up methods needed to be adapted. The transit was fitted using \texttt{Namaste}, and the radial velocity slope was measured and compared to a grid of planet-like orbits with varying masses and periods. These used stellar parameters measured from spectra and the distance as measured by Gaia.}
{Orbiting around a sub-giant star with a radius of 2.70$\pm$0.12R$_{\rm Sol}$, the planet has a radius of 1.11$_{-0.07}^{+0.07}$R$_{\rm Jup}$ and a period of 3650$_{-1130}^{+1280}$ days.
The radial velocity measurements constrain the mass to be lower than 13M$_{\rm Jup}$, which implies a planet-like object.}
{We have found a planet at 4.5 AU from a single-transit event. After a full radial velocity follow-up campaign, if confirmed, it will be the longest-period transiting planet discovered.}


\maketitle

%-------------------------------------------

\section{Introduction}

Detecting exoplanets via single-transit events (monotransits) will be crucial in the era of short-duration (27-day)  campaigns with the \textit{Transiting Exoplanet Survey Satellite} (\textit{TESS),}  with over 1000 monotransits estimated \citep{Villanueva2018}.  To date, several monotransit candidates have been proposed \citep{Osborn2016,LaCourseJacobs2018,Osborn2018,Vanderburg2018}. \citet{LaCourseJacobs2018} listed more than 160 candidates and also reported the detection of the monotransit we study here.
However, only one monotransit has been confirmed and was reobserved (HIP116454b, \citealt{Vanderburg2015}). This transit is on a 9.1-day orbit.

We report the discovery of EPIC248847494b, a sub-stellar object on a very long-period orbit that exhibited a single transit in Campaign 14 of \textit{K2}. In Section~\ref{sec:Obs} we outline the observations that lead to and followed the detection. In Section~\ref{sec:analysis} we describe the analysis of the data we performed to characterise the system, and the processes we used to eliminate possible causes other than a transit. In Section~\ref{sec:disc} we discuss the implications of this planet-like object, and in Section~\ref{sec:conc} we summarize the discovery.

\section{Observations}
\label{sec:Obs}
The source EPIC248847494b was observed in Campaign 14 of the \textit{K2} mission with long-cadence (29.4-minute) exposures. The campaign began on 1 June 2017 at 05:06:29 UTC and ended on 19 August 2017 at 22:11:02 UTC, lasting 79.7~days. 

\begin{figure}
\centering
\includegraphics[width=9cm]{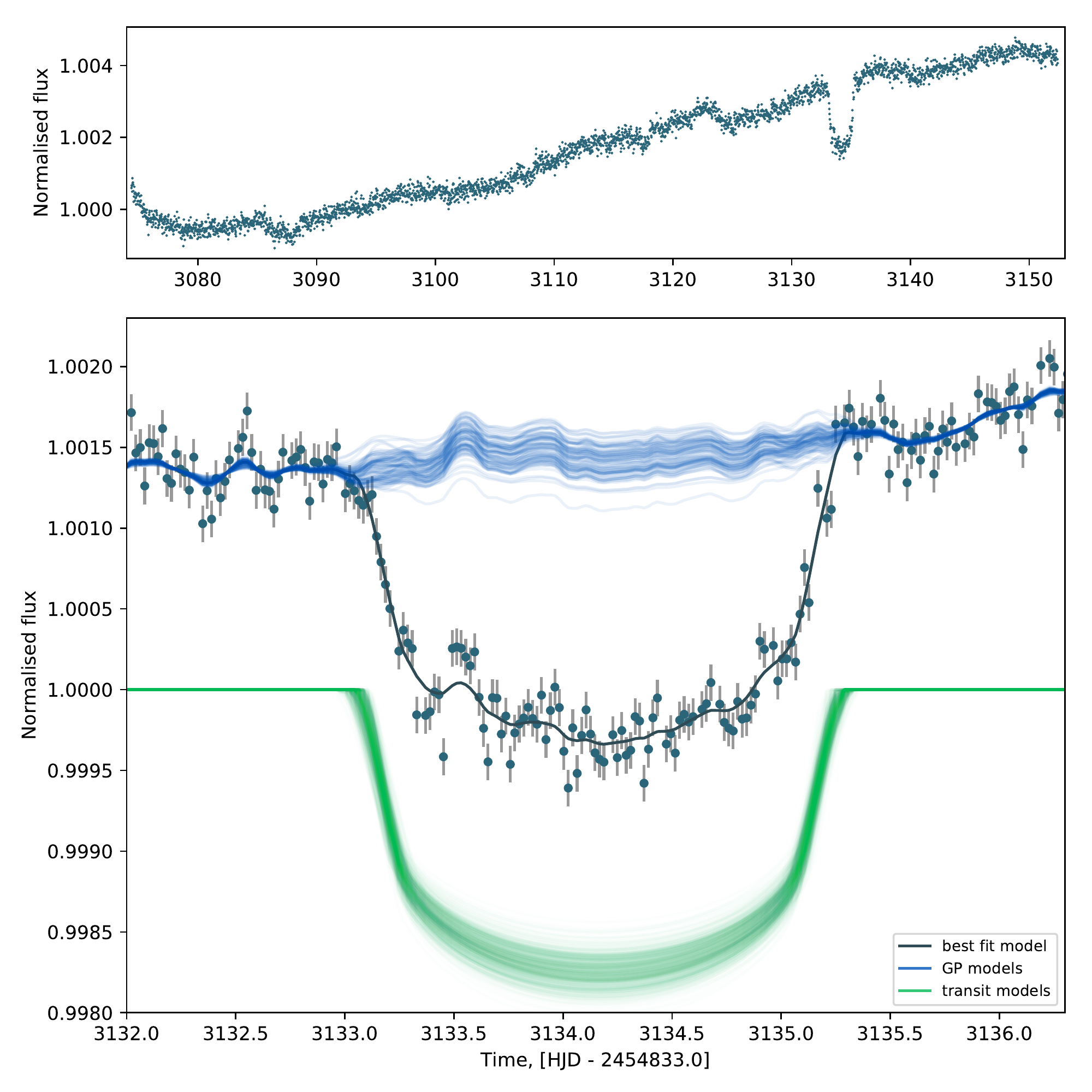}
\caption{Transit of EPIC248847494b observed by \textit{K2} and \texttt{Namaste} models. The upper panel shows the full light curve, and the lower panel shows a zoom of the transit together with the models. The black line shows the best-fit \texttt{Namaste} model. This is composed of the transit model (100 randomly selected models shown in green), and Gaussian process realisations (blue).}
\label{fig:lc}
\end{figure}

Following the public release of \textit{K2} reduced data on 20 November 2017, the light curves were searched for planetary signals following the same method as described in \citet{Giles2018}. This method uses the K2 PDC\_SAP-reduced light curves, which we detrended using a moving polynomial, and we removed significant outliers. Then we searched for transits using a box-fitting least-squares algorithm \citep[BLS,][]{Kovacs2002}. In addition to regular transit candidates, we detected a single-transit event in the light curve of EPIC248847494 (see Fig.~\ref{fig:lc}). The transit depth is approximately 1.7 mmag, lasting over 53~hours. No other transits or unusual systematics were seen in the light curve. From this we conclude that the event is of astrophysical origin.

\begin{figure}
\centering
\includegraphics[width=9cm]{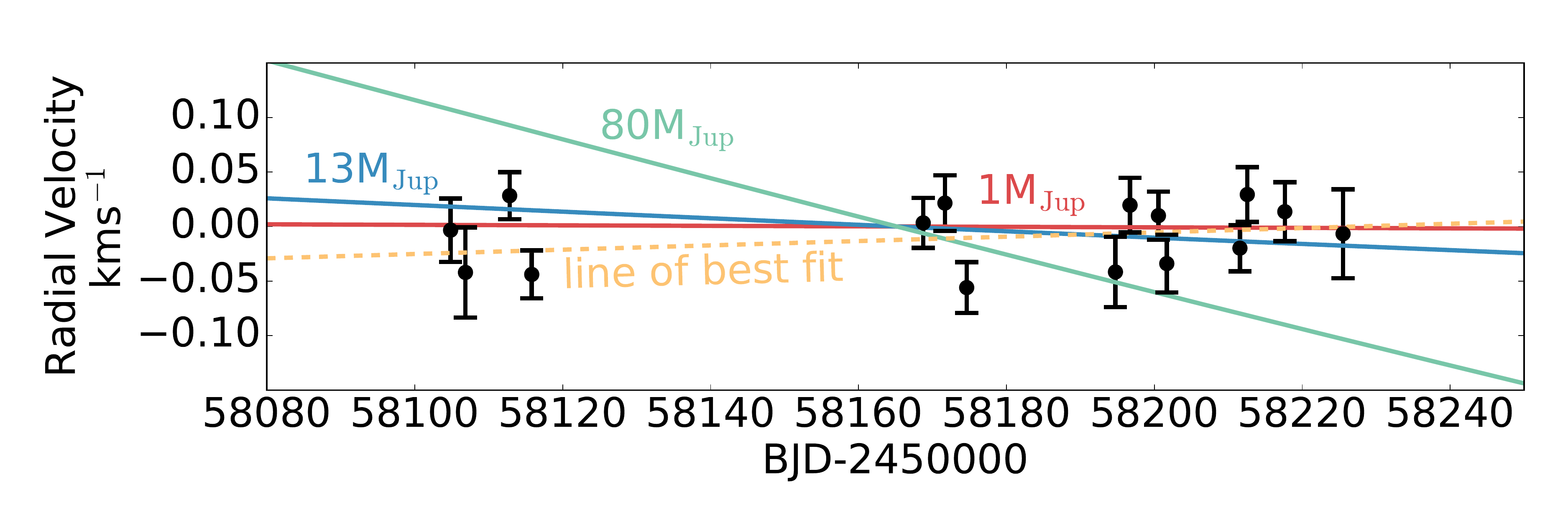}
\caption{Radial velocity observations from CORALIE (black points) compared with circular-orbit models of three objects: a Jupiter-mass planet (red), a 13M$_{\rm Jup}$ brown dwarf (blue), and an 80M$_{\rm Jup}$ low-mass star (green), assuming a period of 3650 days. The yellow dashed line is the best-fit line (see Sect.~\ref{sec:Obs}).}
\label{fig:rvs}
\end{figure}

In order to determine the nature of this very strong candidate, we observed EPIC248847494 with the 1.2m Euler telescope at the La Silla Observatory in Chile using the CORALIE spectrograph \citep{Queloz2000}. CORALIE is a fibre-fed, high-resolution (R=60,000) echelle spectrograph that is capable of high-precision ($<6 \rm ms^{-1}$) radial velocity measurements (RVs). Fifteen observations were taken between 17 December 2017 and 17 April 2018 (see Table~\ref{tab:rv}), where a 16th point was removed because of significantly high instrumental drift. These points give an RV slope of 0.19$\pm$0.16 m s$^{-1}$ day$^{-1}$ (Fig.~\ref{fig:rvs}). 

\begin{table}
\caption{CORALIE radial velocities of EPIC248847494}
\label{tab:rv}
\centering
\begin{tabular}{l l l r}
\hline\hline
BJD-2450000 & RV [$\rm kms^{-1}$] & RV Error [$\rm kms^{-1}$] & BIS\\
\hline
8104.845468     & 29.088 & 0.029 & 0.015\\
8106.856642     & 29.050 & 0.041 & 0.001\\
8112.830676     & 29.120 & 0.022 & -0.061\\
8115.818047     & 29.048 & 0.022 & -0.020\\
8168.748372 & 29.095 & 0.023 & 0.000 \\
8171.685229 & 29.113 & 0.025 & -0.061 \\
8174.628598 & 29.036 & 0.023 & -0.036 \\
8194.743602 & 29.050 & 0.032 & -0.034 \\
8196.710862 & 29.111 & 0.025 & -0.010 \\ 
8200.554695 & 29.102 & 0.022 & -0.004 \\
8201.685863 & 29.058 & 0.027 & -0.001 \\
8211.576046 & 29.072 & 0.021 & -0.011 \\
8212.572145 & 29.121 & 0.025 & -0.055 \\
8217.646377 & 29.105 & 0.027 & -0.037 \\
8225.522768 & 29.085 & 0.041 & -0.038 \\
\hline
\end{tabular}
\end{table}

To check that RV variations were not due to a blended spectrum, we computed the bisector slope of the cross-correlation function for each observation as described by \citet{Queloz2001}, see Table~\ref{tab:rv}. We see no correlation between the bisector slope and radial velocities. We also recomputed this using different stellar masks but found no trends, which suggests that this is not a blended binary \citep{Bouchy2009}.

\section{Analysis}
\label{sec:analysis}
\subsection{Stellar parameters}

\begin{table}
\caption{Properties of the EPIC248847494 system}
\label{tab:K2cand}
\centering
\begin{tabular}{l l r}
\hline\hline
Parameter & Units & Value \\
\hline
\multicolumn{3}{l}{Stellar parameters} \\
2MASS & \multicolumn{2}{r}{J10373341+1150338$^{\rm a}$} \\
$\alpha$ & Right ascension [hh:mm:ss] & 10:37:33.42$^{\rm a}$\\
$\delta$ & Declination [dd:mm:ss] & 11:50:33.8$^{\rm a}$\\
Kep & [mag] & 12.17$^{\rm a}$\\
V & [mag] & 12.42$^{\rm b}$\\
K & [mag] & 10.15$^{\rm c}$\\
g$_{\rm Gaia}$ & [mag] & 12.17$^{\rm d}$\\
$\mu_{\rm \alpha}$ & Proper motion [mas yr$^{\rm -1}$] & -38.74$\pm$0.07$^{\rm d}$\\
$\mu_{\rm \delta}$ & Proper motion [mas yr$^{\rm -1}$] & 1.21$\pm$0.06$^{\rm d}$\\
$\hat{\rm \pi}$ & Parallax [mas] & 1.78$\pm$0.04$^{\rm d}$\\
d & Distance [parsecs] & 560$\pm$13$^\dagger$\\
Fe/H & Metallicity [dex] & -0.23$\pm$0.04$^\dagger$ \\
T$_{\rm eff}$ & Effective temperature [K] & 4898$\pm$68$^\dagger$ \\
log(g) & Surface gravity [dex] & 3.52 (fixed)$^\dagger$ \\
R$_*$ & Radius [R$_{\rm sol}$] & 2.70$\pm$0.12$^\dagger$ \\
M$_*$ & Mass [M$_{\rm sol}$] & 0.90$\pm$0.09$^\dagger$ \\
$\rho_*$ & Density [g cm$^{\rm -3}$] & 0.064$\pm$0.007$^\dagger$ \\
$\mu_1$ &  Lin. limb-darkening coeff. & 0.562$^{-0.001}_{+0.001}$\\
$\mu_2$ &  Quad. limb-darkening coeff. & 0.149$^{-0.001}_{+0.001}$\\
\\
\multicolumn{3}{l}{Planet parameters} \\
P$_{\rm orb}$ & Period [days] & 3650$_{-1130}^{+1280}$$^\dagger$\\
$v'$ & Orbital velocity [R$_{*}$d$^{-1}$] & 0.61$_{-0.05}^{+0.08}$$^\dagger$\\
T$_{\rm C}$ & Transit centre [BJD] & 2457967.17$_{-0.01}^{+0.01}$$^\dagger$\\
T$_{\rm D}$ & Transit duration [hours] & 53.6$_{-5.3}^{+5.9}$$^\dagger$\\
R$_{\rm P}$/R$_{\rm *}$ & Planet-stellar radii ratio & 0.042$_{-0.002}^{+0.002}$$^\dagger$\\
a & Semi-major axis [AU] & 4.5$_{-1.0}^{+1.0}$$^\dagger$\\
b & Impact parameter & 0.79$_{-0.07}^{+0.04}$$^\dagger$\\
i & Inclination [$^{\circ}$] & 89.87$_{-0.03}^{+0.02}$$^\dagger$\\
R$_{\rm P}$ & Planet radius [R$_{\rm Jup}$] & 1.11$_{-0.07}^{+0.07}$$^\dagger$\\
<F> & Incident flux [ergs s$^{-1}$ cm$^{-2}$] & 2.6$_{-0.9}^{+1.7} \times 10^5$$^\dagger$\\
T$_{\rm eq}$ & Equilibrium temperature [K] & 183$_{-18}^{+25}$$^\dagger$\\
\hline
\multicolumn{3}{l}{$^{\rm a}$ \citealt{Huber2016}, $^{\rm b}$ APASS: \citealt{APASS},}\\
\multicolumn{3}{l}{$^{\rm c}$ 2MASS: \citealt{2MASS},}\\
\multicolumn{3}{l}{$^{\rm d}$ \citealt{Gaia2018}, $^\dagger$ This Work}\\
\end{tabular}
\end{table}

To determine the stellar parameters of EPIC248847494, we followed the same method as \citet{Giles2018}. A pipeline was built for the CORALIE spectra based on \texttt{iSpec}\footnote{\url{http://www.blancocuaresma.com/s/iSpec}} \citep{iSpec}.  All observations were aligned and co-added to increase the signal-to-noise ratio (S/N), were reduced and spectrally fitted using the code SPECTRUM \citep{Gray1994} as the radiative transfer code. Atomic data were obtained from the Gaia-ESO Survey line list \citep{Heiter2015a}. We selected the line based on an $R\sim47\,000$ solar spectrum \citep{Blanco2016, Blanco2017}, and we used MARCS model atmospheres \citep{Gustafsson2008}. The resulting errors were increased by quadratically adding the dispersions found when analysing the Gaia benchmark stars \citep{Heiter2015b, Jofre2014, Blanco2014} with the same pipeline.  This resulted in an effective temperature of 4877$\pm$68K, a log g of 3.41$\pm$0.07 dex, and [Fe/H] = -0.24$\pm$0.04 dex.

In the second data release of Gaia \citep{Gaia2018}, EPIC248847494 has a measured parallax (see Table~\ref{tab:K2cand}) based on which we can determine an independent stellar radius using bolometric absolute magnitudes and the spectroscopically determined effective temperature for EPIC248847494 following the method detailed in \citet{FultonPetigura2018}. We took the K-band apparent magnitude \citep{2MASS}, the Gaia distance, and a bolometric correction (BC$_{\rm K}$, from \citealt{Houdashelt2000}) of 1.91$\pm$0.05, which was interpolated from the range within the coarse grid. We chose not to include an extinction correction as this only introduces an uncertainty of ~0.5\% \citep{FultonPetigura2018}. This gave a radius of $2.70\pm0.12 {\rm R_{sol}}$.

Taking the spectrally determined metallicity and effective temperature and the measured radius as observational constraints, we input them into the Geneva stellar evolution code \citep{Eggenberger2008}. This resulted in a stellar mass of 0.9${\pm 0.09 \rm M_{sol}}$.
These values of mass and radius would therefore indicate a log g of 3.52 dex. When we fixed the iSpec analysis to this log g, the metallicity and effective temperature were very similar to the initial results (see Table~\ref{tab:K2cand}). Log g is not well constrained spectroscopically, and changes have a very limited effect on other parameters. Therefore we adopt the parameters based on log g = 3.52.

\subsection{Eliminating the photometric systematics of K2}
\label{sec:eliminate}
The possibility for false positives is high in monotransits. We therefore endeavored to eliminate all causes for false positives.  All objects listed as `stars' with \textit{K2} light curves within 25 arcminutes were checked for similar artefacts. Of the 61 objects, none showed odd behaviour at the same epoch as the monotransit. Additionally, the location of EPIC248847494 was not near the edge of the CCD, which suggests that no near-edge effects occurred.  In the target pixel file of EPIC248847494, we checked the pixels for changes and failures before (both the star and background flux), during, and after the transit, but found none.  We checked the centroid shifts of EPIC248847494 in the \textit{K2} release light curves. Pointing has three clear regimes (times given in BJD-2454833): $\sim$3072-3087 days, which is when \textit{K2} settles into position after changing field; $\sim$3087-3124 days, which is when \textit{K2} approaches optimum stability position; and $\sim$3124-3153 days, when \textit{K2} leaves the optimum stability position.  The optimum stability position is the moment when the balance between the remaining reaction wheels of \textit{K2} is most stably balanced against the solar radiation pressure (G. Baretsen, personal communication).
The monotransit is away from this optimum stability position and other shifts in pointing. Furthermore, there is no evidence that the centroid position for the point spread functions (PSFs) or the flux-weighted centre have dramatically changed for any reason. Using the extracted light curve from \citet{Vanderburg2014}, which is available from MAST\footnote{\url{https://archive.stsci.edu/prepds/k2sff/}}, we checked the in-transit points along the measured arc caused by the movement of \textit{K2}. When we inspected the change in flux that is due to arclength, no in-transit points were constrained to a single area, but the points covered the arc uniformly with no evidence for earlier or later points favouring certain arclength positions.  No close neighbours are present in the Gaia DR2 data \citep{Gaia2018}.

\subsection{Planet parameters}
\label{sec:planetparams}

General transit-fitting methods are often not suitable for the modelling of monotransits, as intrinsic knowledge of the orbit is necessary (e.g. $P$ and $R_*/a$), therefore a monotransit-specific fitting code \citep[\texttt{Namaste},][]{Osborn2016}\footnote{\url{https://github.com/hposborn/namaste}} was used to model the HLSP light curve from \citet{Vanderburg2014} of EPIC248847494 and explore the planetary characteristics.
The code applies the transit models of \citet{MandelAgol2002}, taking the lateral velocity of the planet (scaled to stellar radius) as a parameter.
Other transit parameters required are planet-to-star radius (uniform prior between 0.02 and 0.25), impact parameter (uniform prior between $-1.2$ and 1.2), transit centre, and limb darkening.
Quadratic limb-darkening coefficients were estimated from T$_{\rm eff}$, log g, and metallicity for the \textit{Kepler} bandpass \citep{Sing2010} and were fixed using a Gaussian prior.

The code\texttt{\textup{} Emcee} \citep{ForemanMackey2016} was used to explore the parameter space of the transit and Gaussian process (GP) models.
To model the stellar and photon noise in the light curve, we used the \texttt{celerite} Gaussian process package \citep{celerite}. We fit two GPs, an exponential kernel for long-timescale trends ($\log{a}= -7.41 \pm 0.72$, $\log{c} = -10.9 \pm 0.7$), and a matern-3/2 kernel for short-timescale granulation ($\log{\sigma} = -10.0^{+0.23}_{-1.3}$, $\log{\rho}= -2.72^{+0.48}_{-0.38}$), alongside a fixed white-noise term (90ppm, \citealt{KeplerHandbook}). We also tested the performance of a stellar rotation-like quasi-periodic kernel, an artificially high white-noise term to account for granulation, and fitting rather than fixing the white noise, all of which gave consistent results.

The best fit is a planet with an orbital velocity of $v'$=0.61$_{-0.05}^{+0.08}$ R$_*$ d$^{-1}$ , which gives an orbital period of 3650$_{-1130}^{+1280}$ days when converted using
\begin{equation}
\left(\frac{P_{\textrm{circ}}}{d}\right)=18226\frac{\left(\rho_{\star}/\rho_{\odot}\right)}{\left(v'/d^{-1}\right)^{3}}.
\end{equation}
However, the model fitting revealed strong correlations between R$_{\rm P}/\rm R_{\rm *}$, b, and $v'$.
This suggests that a slightly smaller planet with high velocity on a low-impact parameter transit fits the data almost as well as the larger R$_{\rm P}/R_{\rm *}$ and $b$ but lower $v'$.
Because R$_{\rm P}$/R$_{\rm *}$ only varied by a small amount, it did not significantly change the planetary radius.

The \texttt{Namaste} fit resulted in a planet-like object with a radius of 1.11$\pm0.07$R$_{\rm Jup}$, orbiting its host star between 3.5 and 5.5 AU. This would indicate the planet has a temperature of approximately 183$_{-18}^{+25}$ K (with the albedo set to 0).
For simplicity, we assumed an eccentricity of 0, although we note that any orbital eccentricity would increase the spread on the velocity and therefore the period. For details, we refer to \citet{Osborn2016}. We hope to constrain this as we gather more long-term RV data.

\begin{figure*}
\centering
\includegraphics[width=\textwidth]{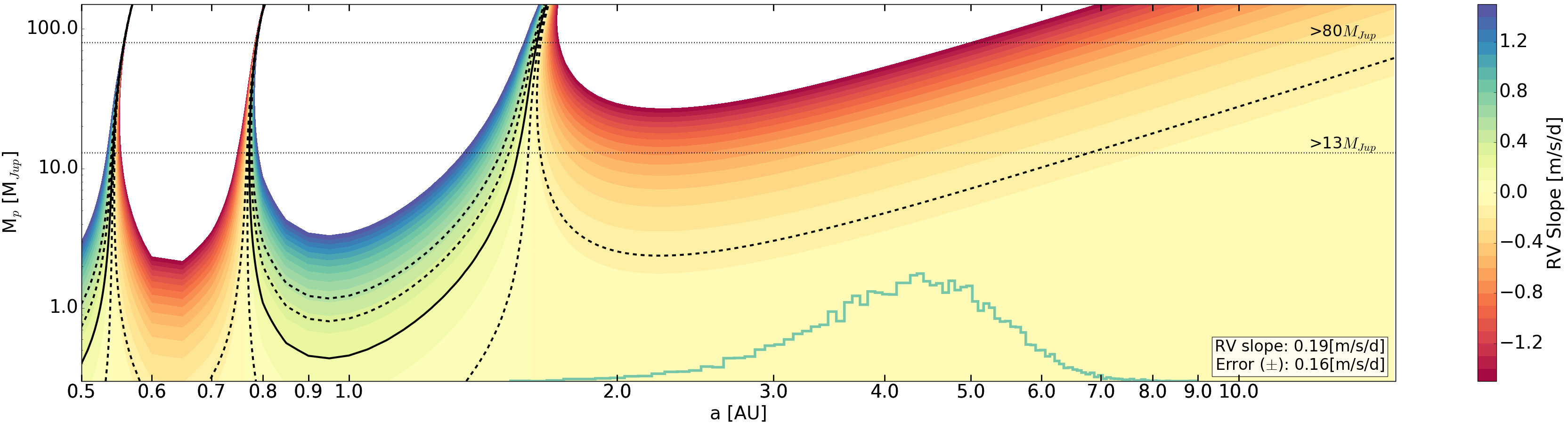}
\caption{Grid of semi-major axes and planetary masses and their corresponding RV slopes, using observations from CORALIE. The colour scale ranges -1.5 to 1.5 m/s/day, with all else set to white. The slope, with 1 and 2$\sigma$ errors, of the CORALIE RVs (solid and dashed black lines) shows regions of likely solutions. We also show mass limits for low-mass stars and brown dwarfs (black dashed lines). The \texttt{Namaste} fit of the light curve (see Sect.~\ref{sec:analysis}) produces a distribution of semi-major axes (green histogram).
The peaks in the grid scale at 0.55, 0.75, and 2 AU are due to RV quadrature for these orbits.}
\label{fig:grid}
\end{figure*}

Knowing the time of transit means that we are in a unique position for RV follow-up. For all observations, it is possible to calculate the phase given an orbital period or semi-major axis, and an RV value given a planetary mass. Therefore we constructed a grid of semi-major axes, 0.5 to 15AU, and planetary masses, 0.3 to 150M$_{\rm Jup}$. Based on this, we calculated the orbital period and the semi-amplitude for the system, assuming that the eccentricity is zero.
We calculated for each grid point the RVs that would occur at the times for which we have data and determined the RV slope, assuming a linear fit, in m/s/day. In Figure~\ref{fig:grid} we show the measured RV slope and the 1 and 2$\sigma$ errors that cover the estimated semi-major axis range from \texttt{Namaste}. The peaks in the grid scale at 0.55, 0.75, and 2 AU are due to RV quadrature for these orbits. In combination with Fig.~\ref{fig:rvs}, it is clear that the RV signal would indicate a mass of 13M$_{\rm Jup}$ or lower.

We also calculated the minimum RV slope we would expect to see for certain celestial body types in the 4.5 AU orbit from \texttt{Namaste}. For a low-mass star (>80M$_{\rm Jup}$) and a brown dwarf (>13M$_{\rm Jup}$), we would expect to see 1.88 m/s/day and 0.31 m/s/day, respectively. Therefore a planet-like object would be required to show a change over $\sim120$ days of less than $\sim$36m/s (Fig.~\ref{fig:rvs}).

\section{Discussion}
\label{sec:disc}

If EPIC248847494b is indeed planetary in nature and confirmed with RVs, it will be the transiting exoplanet with the longest ever discovered period. A final confirmation would require three years of RV follow-up.
Currently, there is only one confirmed transiting planet in the NASA Exoplanet Archive\footnote{\url{exoplanetarchive.ipac.caltech.edu}} \citep{Akeson2013} with a period longer than 2500 days (our lower limit).
With an occurrence rate of $\sim$4.2\% \citep{Cumming2008} for a planet with mass between 0.3 and 15 M$_{\rm Jup}$ in a 3-6 AU orbit and a transit probability of 0.12\%, applied to the entire \textit{K2} catalogue (312,269 stars) that is observed for a maximum of 80 days, we would expect to detect about one object.

Based on a comparison with planets within the solar system, EPIC248847494b is similar to our gas giants, which strongly suggests that it possesses moons. The estimated equilibrium temperature of 183$_{-18}^{+25}$K would indicate that the planet is close to the snow line. Therefore, any moons may well be near the habitable zone, based on the stellar effective temperature and luminosity \citep{Kopparapu2013,Kopparapu2014}, although it would have been much cooler for most of the main-sequence lifetime of this star. 

The minimum observing windows for \textit{TESS} are 27.4 days (assuming non-consecutive observing windows). This will apply a hard limit of $\sim$28-day periods for objects to have two or more transits.
This has recently been investigated by \citet{Villanueva2018}, who estimated that \textit{TESS} will discover 241 monotransits from the postage stamps and a further 977 from the full-frame images. With the possibility of over 1000 new single-transit candidates, there may be many more EPIC248847494b-type planets to be discovered and characterised.

\section{Conclusions}
\label{sec:conc}

In Campaign 14 of the \textit{K2} mission, we detected a monotransit in the light curve of EPIC248847494 and performed follow-up observations. Based on the spectra we obtained as RV measurements, we determined that EPIC248847494b orbits a $2.70\pm0.12$ R$_{\rm sol}$ star with a mass of 0.9$\pm 0.09$ M$_{\rm sol}$ , that is, a sub-giant star.
EPIC248847494b is the first long-period planet to be vetted using RV, starting from a single monotransit. We estimate the orbital period to be 3650$_{-1130}^{+1280}$ days, the radius to be approximately 1.11$\pm0.07$ R$_{\rm Jup}$ , and we derive a lower and upper limit on the mass of 1 and $\sim$13 M$_{\rm Jup}$ , respectively.

This is an excellent candidate for which to  attempt detecting exomoons that may well be habitable. This would require extremely precise photometry (e.g. CHEOPS, \citealt{CHEOPS}, or PLATO, \citealt{PLATO}) for future transit events, however.

Additionally, given the shorter observation campaigns of \textit{TESS}, the number of monotransit candidates will increase. We have shown that it is possible, given the parameters that can be measured from the transit, to characterise these candidates and potentially push detections to increasingly longer orbital periods.

\begin{acknowledgements}
We would like to thank the referee for their comments and suggestions that were extremely valuable for the development of this Letter.\\
We thank the Swiss National Science Foundation (SNSF) and the Geneva University for their continuous support of our planet search programs. This work has  specifically been carried out in the frame of the National Centre for Competence in Research `PlanetS’ supported by the Swiss National Science Foundation (SNSF).\\
This paper includes data collected by the \textit{Kepler} mission. Funding for the \textit{Kepler} mission is provided by the NASA Science Mission directorate.\\
HG acknowledges the assistance from L. Temple on measuring stellar radii.\\

\end{acknowledgements}

%-------------------------------------------------------------------

\bibliographystyle{aa} % style aa.bst
\bibliography{references}

\end{document}